\RequirePackage{fix-cm}
\documentclass[smallextended]{svjour3}       
\smartqed  
\usepackage{graphicx}
\usepackage[small]{caption}

\newcommand{\rem}[1]{}

\begin{document}

\title{Kenneth G. Wilson:\\ Renormalized After-Dinner Anecdotes}

\author{Paul Ginsparg}

\institute{P. Ginsparg \at
              452 Physical Sciences Building\\
              Cornell University\\
              Ithaca, NY 14853\\
              \email{ginsparg@cornell.edu}           
}

\date{}  

\maketitle
\vskip-50pt

\begin{abstract}
This is the transcript of the after-dinner talk I gave at the close of the 16 Nov 2013 symposium ``Celebrating the Science of Kenneth Geddes Wilson'' \cite{kws} at Cornell University (see Fig.~\ref{fig:0} for the poster).  The video of my talk is on-line \cite{kwpg}, and this transcript is more or less verbatim,\footnote{I was unable to prepare detailed notes in advance of the talk, so instead transcribed this text more than a half year later from the video of a largely extemporaneous presentation.  I retained the run-on sentences for verisimilitude,
but deleted multiple instances of the word `So $\ldots$',  which preceding virtually every other sentence appeared too distracting in print.  The footnotes are later additions.}
with the slides used included as figures. I've also annotated it with a few clarifying footnotes, and provided references to the source materials where available.


\end{abstract}

\section{Introduction}
\label{intro}

\begin{figure}
\includegraphics[width=\textwidth]{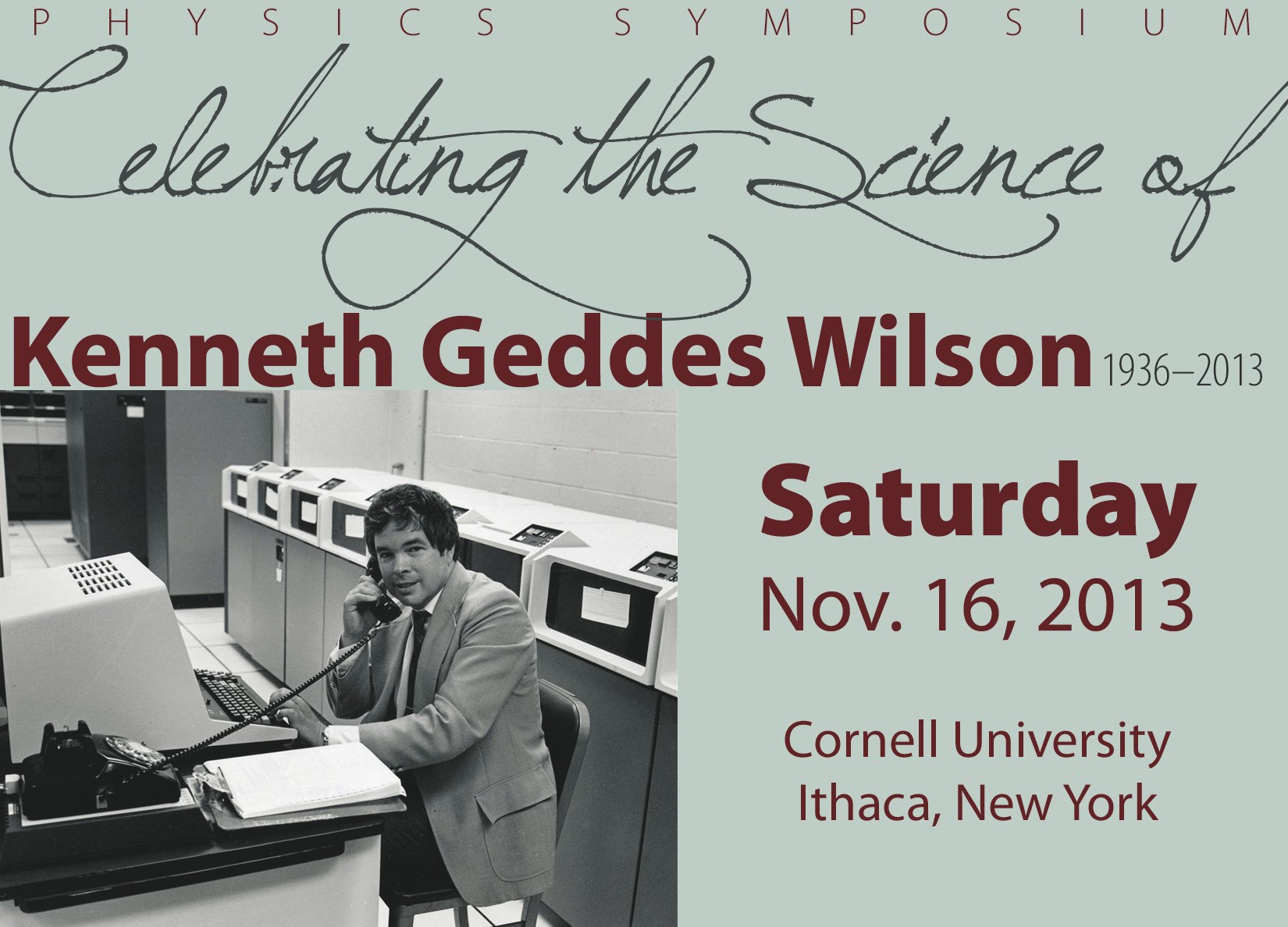}
\caption{Symposium Poster}
\label{fig:0}
\end{figure}

I've been invited to give the after-dinner talk to complete today's celebration of Ken Wilson's career.
It is of course a great honor to commemorate one's  thesis advisor  on such an occasion --- and a unique opportunity.
Earlier today, Ken's wife, Alison (sitting over here), asked me to be irreverent tonight. (``Don't blame it on me," she calls out. ``Not hard for you to do," says someone else.)
Actually I replied, ``You've got the wrong person, I only do serious.''
Since the afternoon events ran a bit late, there wasn't quite as much time as I'd hoped to absorb the events of the day and think about what to say tonight.  But by the standards of this afternoon, I hope you're all prepared to be here until midnight $\ldots$ 

As Ken's student, I had only a brief window of interaction with him.
Many of his colleagues here knew him from 1963 to 1988 at Cornell, 
or, like David Mermin, knew him since they were undergrads together starting in 1952.
I only had two or three years of interaction with him as a graduate student,\footnote{I was a Cornell graduate student from 1977--1981. Ken was chair of my committee starting in 1978. During the academic year 1979--1980, he was on sabbatical in Zurich, and I spent the year at C.E.N.\ Saclay, outside of Paris; but we met a few times that year. We wrote one article \cite{GW} together during my final year, related to my thesis work.}
after he became the chair of my special committee.
But nonetheless as far as I can tell I've retained many more anecdotes about him.
When I'd asked the people who knew him for decades,
``Do you have any interesting stories?" They'd say simply, ``No.''
``Well did you go hiking with him?''  ``Yes, I went hiking with him.''
``Well did he say anything then?''  ``No, he was very quiet'' $\ldots$

I suspect we all have some special place in our memory for  interactions with our thesis advisor.
In Ken's case, of course this was very much amplified because 
we were all sort of in awe of him (as Steve White also mentioned this afternoon).
It's not because there was anything special about him personally, just the opposite: he looked and acted like 
a graduate student, had his bicycle in his office, would crack lame jokes, looked bored in seminars, etc. But we were all aware that this normal-looking guy had transformed the understanding of quantum field theory, of statistical mechanics, and so on.
And apparently unlike his brother David,\footnote{His brother David Wilson, a faculty member at Cornell in Biology, had commented during the afternoon session that he'd been entirely unaware that a trip to Stockholm was inevitable.} we all knew this mild-mannered, unassuming figure was certain to win a Nobel prize.\footnote{I remember struggling to decide whether to say something further about the
dramatic mismatch between Ken's outsized reputation within the physics community and his lack of name recognition from without.
But I didn't really know how to formulate the reasons, and still don't.
It could be that his work was so fundamental and far-reaching that it's impossible to encapsulate in a few short sound-bites;
or that so much of his creative work was concentrated in a single decade from the mid 60's to mid 70's, and he was not as influential in his later career projects;
or that he had little interest in writing popularizations (see, e.g., excerpt \#13 of the video transcripts) or self-promotion.
In short, he probably didn't care.}

So in preparation for this talk, in the last couple of evenings,
I've read various historical materials. I will draw  from
things he's written such as his Nobel lecture \cite{nobel},
an oral history he did with the Dibner Institute \cite{dibner1,dibner3},
and also some videos \cite{videos,excerpts} of him that I've just watched.
And I'll try to tell something of a story,  putting these together,  superficially discussing various periods in his career, and threading in some of the things  he said directly to me for additional illustration.

\begin{figure}
\includegraphics[width=\textwidth]{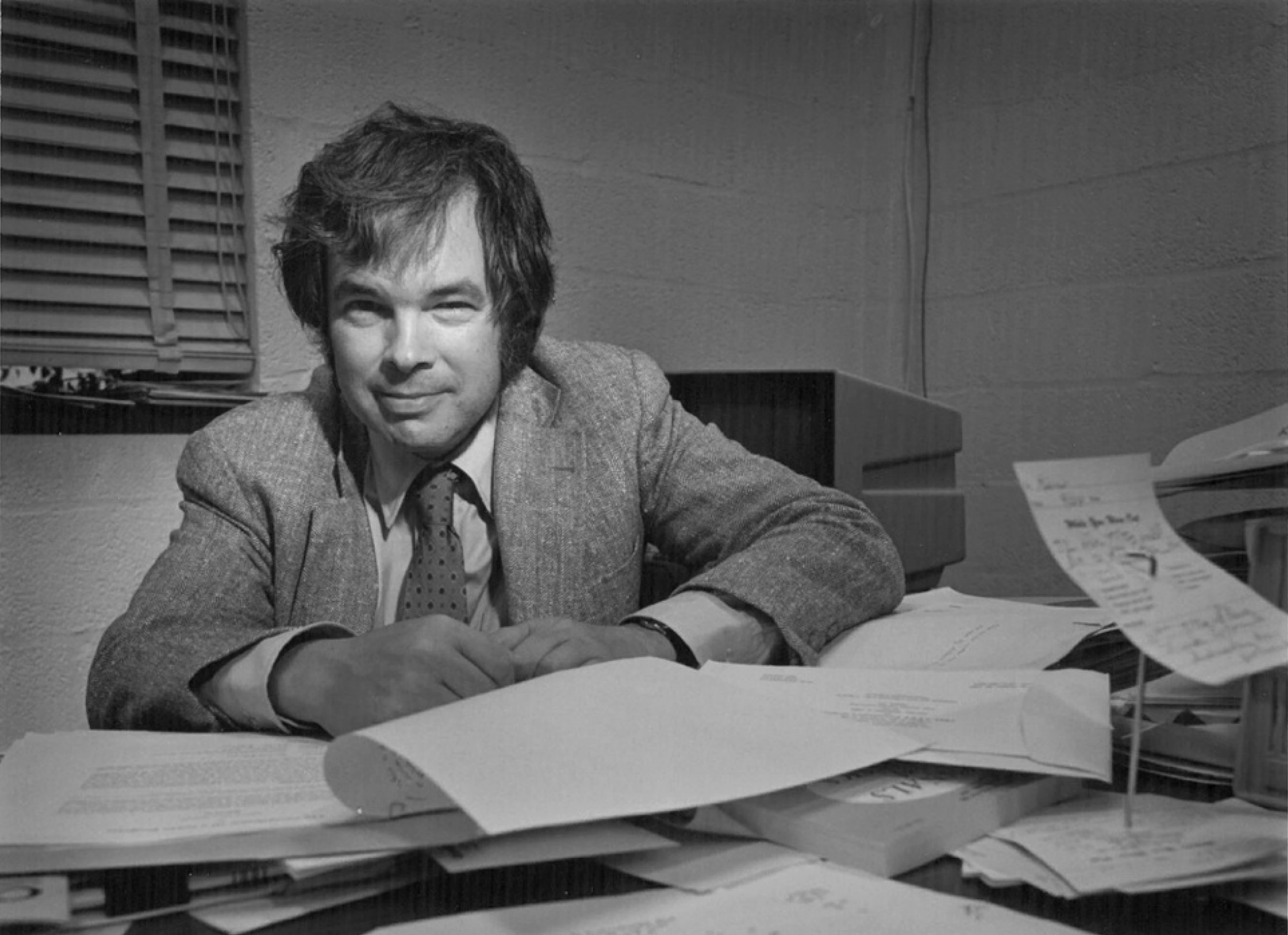}
\caption{Busy at work (before the advent of the paperless office)}
\label{fig:1}   
\end{figure}

To set the stage, here is a photo (Fig.~\ref{fig:1}) more or less along the lines of how we knew him.
It is one of the Nobel promo photos. Amusingly, in so many of these photos we have from that period he's seen
wearing a jacket and tie, but of course he rarely wore a jacket and tie in real life $\ldots$


\begin{figure}[b]
\centering
\includegraphics[width=.75\textwidth]{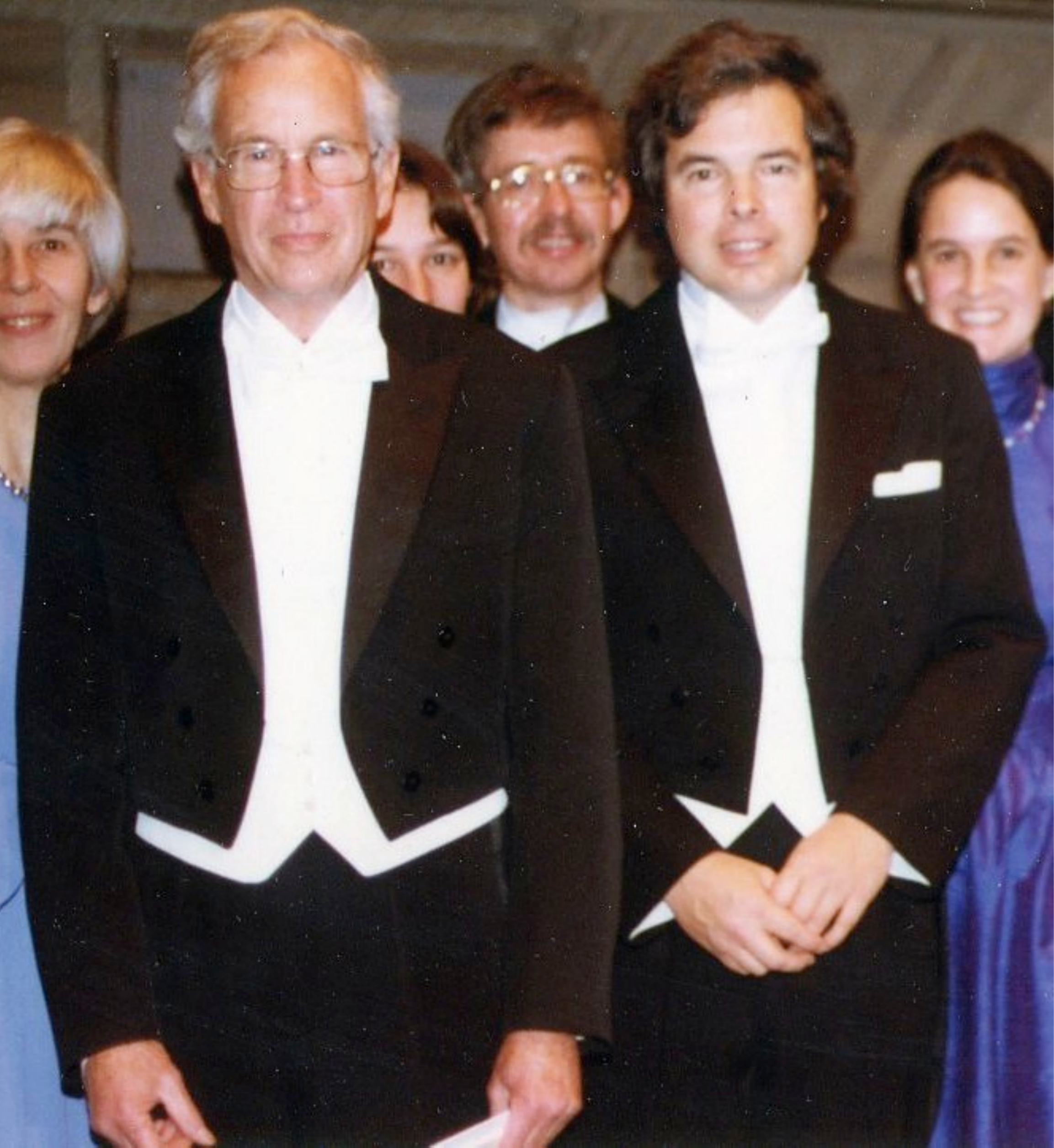}
\caption{Stockholm 1982: Some people really know how to wear white tie and tails}
\label{fig:2}
\end{figure}

\section{Caltech}
\label{caltech}

I'll start with his period at Caltech, and in particular the story of why he became Gell-Mann's  
and not Feynman's student.
To properly explain this story, I have to digress slightly and recall something that was mentioned a few times this morning,  which is that Ken's father, E. Bright Wilson, Jr., was also a very prominent academic,
and a pioneer in the theoretical  and experimental study of the structure of molecules.
Like Ken, he was a former Junior Fellow in the Society of Fellows at Harvard,\footnote{According to Diana Morse at the Harvard Society of Fellows, ``They are the first and only father/son pair both being JFs.  E. Bright's dates as a JF were 1934--36; Ken's dates were 1959--62.''}
 and in 1935 he co-authored with Linus Pauling a  textbook ``Introduction to Quantum Mechanics" \cite{PW}, which was for many decades the standard reference text for physical chemists.

This photo (Fig.~\ref{fig:2}) shows Ken with his father at the Nobel prize event.
I'll always remember something that Ken said to me regarding his father, the Harvard chem prof:
``When I was growing up, I didn't realize that not everyone had a parent
who came home every evening to dinner to fulminate against the stupidity of institutions
and the people who run them."

\begin{figure}[h]
\includegraphics[width=\textwidth]{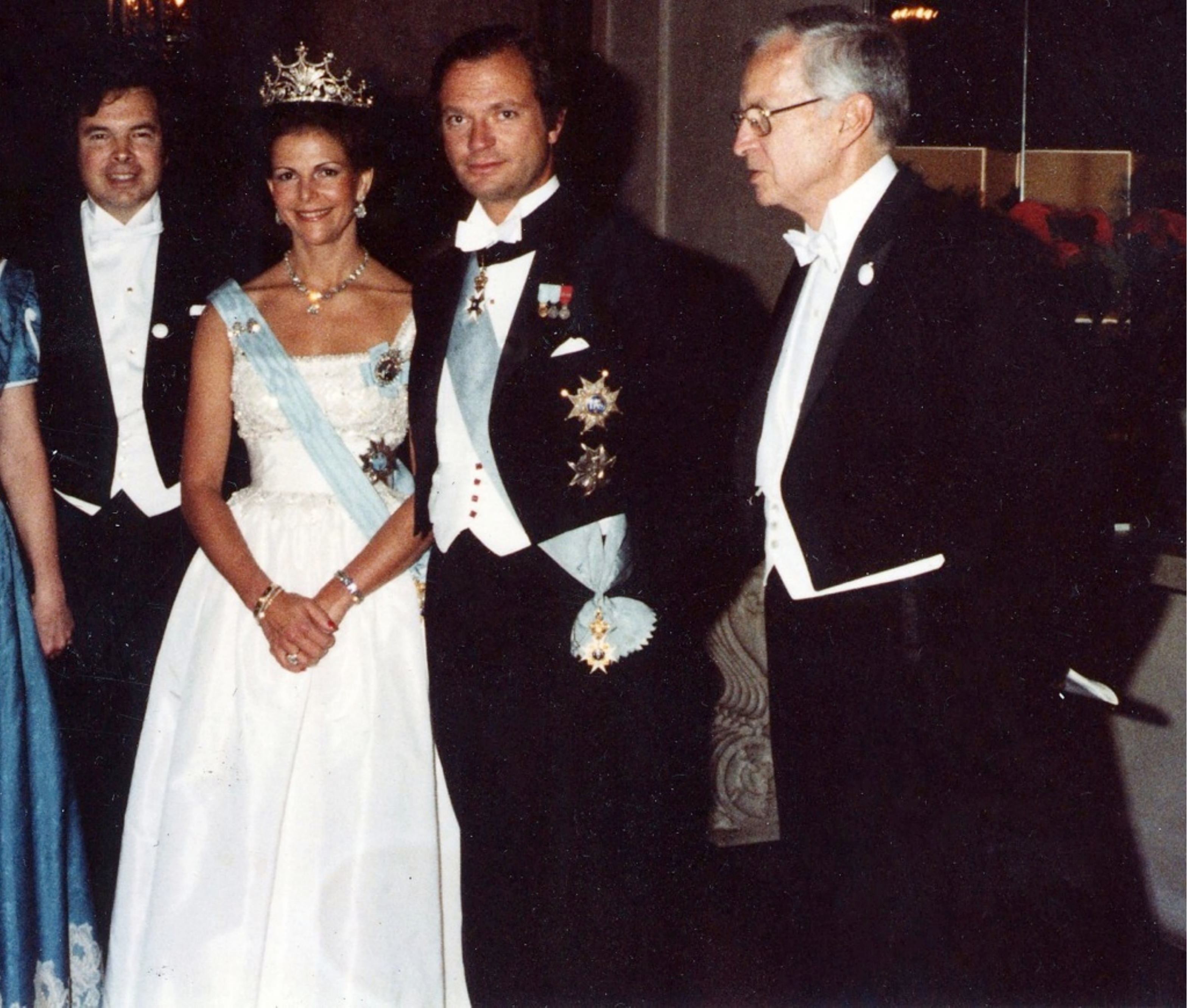}
\caption{``So tell me, is `king' some sort of administrative position at your institution?''}
\label{fig:3}
\end{figure}

We'll see later, in the video excerpts, many positive things Ken says about his father.\footnote{E.g., the transcript of excerpt \#4 in sec.~\ref{videos}}
By a happy circumstance, I also knew his father when I was back at Harvard in the '80s, after getting my doctorate.
Bright Wilson and I were both affiliates of North House,\footnote{An undergraduate residential house at Harvard} so I'd introduced myself and he was  congenial.
It was  slightly surreal I have to say --- I always felt like I was talking to Ken, there was
something eerily reminiscent about his speaking cadence and  use of language.
Here's  another photo of him (Fig.~\ref{fig:3}), no doubt helpfully opining about the institution of monarchy.

His father had instructed Ken when he went to graduate school at Caltech to introduce himself to physicists there. As Ken  wrote \cite{nobel}:
``When I entered graduate school, I had carried out the instructions given to
me by my father and had knocked on both Murray Gell-Mann's and Feynman's
doors, and asked them what they were currently doing."
The sequel to this was told to me by Pierre Ramond.\footnote{I spoke to Pierre about Ken in Aspen this past summer (July 2013). They met when Ken gave a talk at Syracuse University in 1967, where 
Pierre was a physics graduate student who followed little of the talk.
Coming from an undergraduate engineering background,  Pierre said he also never really understood momentum space, but at least Ken was working in position space, so apparently they hit it off and kept in touch through the years.}
Ken related to Pierre that he'd knocked on Feynman's closed door,
and heard a gruff voice say, ``What do you want?"
Eventually Feynman opened the door, and Ken introduced himself, 
``Hi, I'm a new graduate student here, and just wanted to find out what you're working on?"
And Feynman shouted, ``Nothing!", and  slammed the door.

So Ken went next to Gell-Mann,  asked what he was working on, and as Ken wrote \cite{nobel}:
``Murray wrote down the partition function for the three dimensional Ising model
and said it would be nice if I could solve it.''\footnote{The one-dimensional Ising model was solved by Ernst Ising in 1924. The two-dimensional Ising model was solved by Lars Onsager in 1944. The three-dimensional Ising model has still not been solved exactly.  Ken also refers to this problem in the transcript of video excerpt \#7 in sec.~\ref{videos}.}
And that was the beginning of a relationship.

\begin{figure}[!h]
\includegraphics[width=\textwidth]{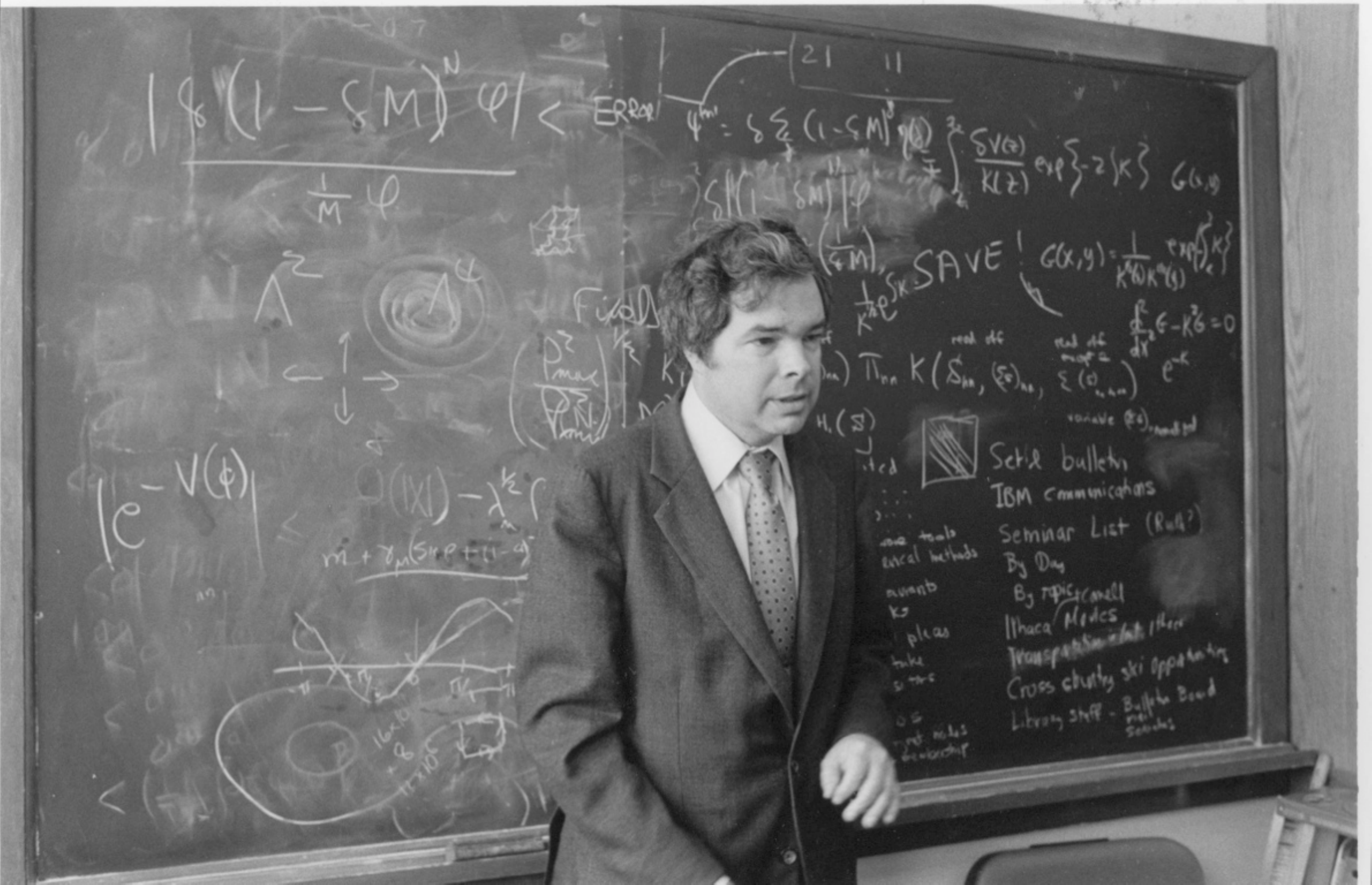}
\caption{Office blackboard (handwriting to the right was later claimed by Peter Lepage)}
\label{fig:4}
\end{figure}

I didn't hear those stories directly from Ken, but for some reason he did once try to illustrate to me
what was for him the take-away from Caltech, making it a uniquely weird place.
He drew for me on the blackboard a series of concentric circles, sort of as in this photo I found (Fig.~\ref{fig:4}, to the left above his shoulder). This isn't the actual one he drew, unless his blackboard went four years without being cleaned. He described it as: ``This is graffiti I saw in a men's room at Caltech. Underneath it,
the caption said, ``photon: full frontal view".
That was about as risqu\'e as it got with him.

\begin{figure}[!h]
\centering
\includegraphics[width=.75\textwidth]{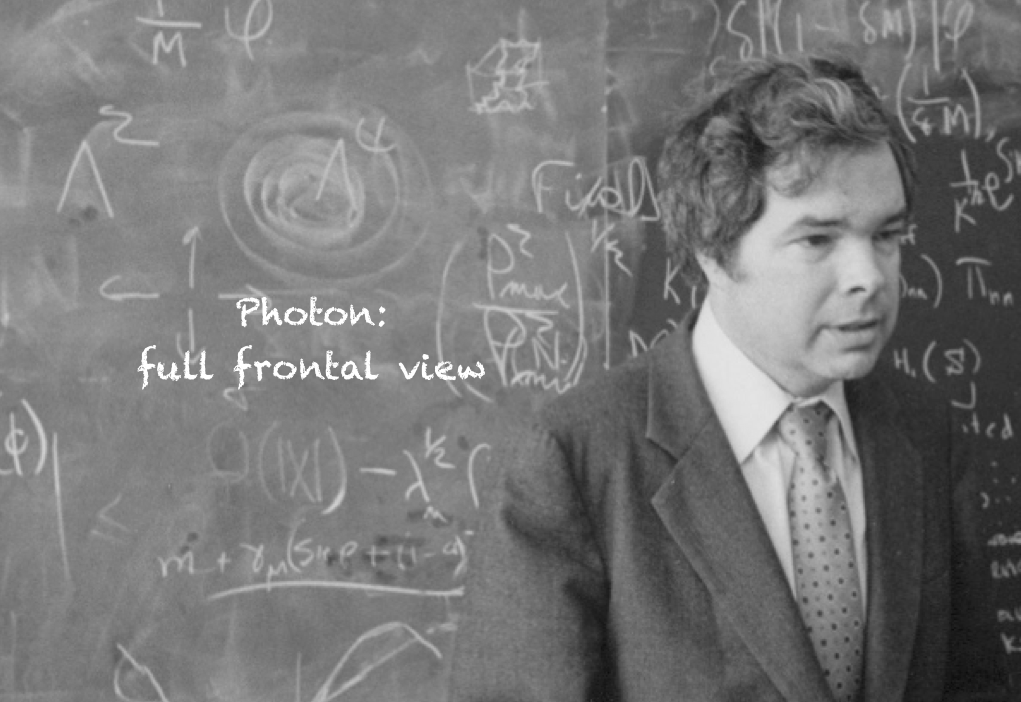}
\caption{Caltech men's room graffiti}
\label{fig:5} 
\end{figure}

\section{Cornell}
\label{cornell}

Moving on to his time at Cornell, there was discussion this morning about the tenure process here for Ken, and whether there was concern over his small number of publications.
Some of you here now were involved in that tenure vote in the 1965 timeframe, 
and everyone I've asked has said, ``No, it was never an issue."\footnote{Explaining the lack of doubt on the part of Ken's colleagues about granting tenure, Vinay A.\ recalls: ``It was obvious from spare comments in seminars and conversations that he {\it understood\/}.''}
Following a discussion once with Vinay Ambegaokar, I  tried to find the actual number of publications back then and counted {\it at least\/} three \cite{f3} that he had written before coming to Cornell in 1963, but there was only one article \cite{t1} that would have counted as a publication from {\it during\/} his first two years at Cornell.\footnote{From the audience, Michael Peskin pointed out that Ken had also worked on the multiperipheral model during that period, and kept returning to it at later times in his career.}
 Evidently referring to that publication, Vinay told me his comment
at the meeting was something along the lines of, ``Too bad he spoiled an otherwise perfect record by publishing an article."
Ken himself, who would not have been present at this meeting (his tenure evaluation), related something complementary \cite{dibner1}: ``Francis Low complained that I should have made sure there was none.  Just to prove that it was possible."

In Aspen this past summer, I also spoke to Steve Berry, a chemist at the University of Chicago, who happened to have been a graduate T.A.\ in a course Ken took as a freshman at Harvard. As a chemist there, Steve also came to know Ken's father, kept in touch with him, and during that period says Bright Wilson had expressed his concern that Ken was working on such obscure stuff
that he would not be able to get tenure. 
So no one was worried, except his father.\footnote{I was told by someone at the meeting that his father later described it as ``a courageus decision'' on Cornell's part.
Pierre Ramond told me that he saw Ken at the Aspen Center for Physics one summer, probably 1970, busily writing up eight articles (appropriately enough) in Bethe Hall. 
Within a couple of years, Pierre says Ken said to him, with a twinkle in his eye,
``You know, I think I'm going to get more citations than my father."}

Ken's father's concern persisted over time. 
Shortly after Ken won the Nobel prize, I was at a North House function,
 Bright Wilson spontaneously came up to me, and without preamble said,
``You know, I think Ken's going through a mid-life crisis."

\section{Thesis advisor}
\label{grad}

Now I turn to some of my experiences as a graduate student here.

I had Ken as lecturer for a wonderful topics course in gauge theories.\footnote{This would have been
the spring of 1979, and the course was audited by many of the postdocs and visitors, and some faculty members as well.
During the course, he calculated the QCD Beta function for $n$ flavors of quarks, and told us
that it was the first time he'd finally gotten around to doing the perturbative calculation.
He'd also said that years earlier he hadn't even wanted to learn the non-abelian group theory, finding the manipulations of structure constants $f_{abc}$ too tedious, until he understood it was essential for understanding the strong interactions.}
(There's at least one other person here, Gyan Bhanot, with whom I remember taking this course;
Serge Rudaz is also here but he didn't seem to go to many courses by then.)
I also audited a graduate course he taught in quantum field theory, where he 
explained all of renormalization in terms of $\lambda \phi^4$ theory. At one point, he broke things up into
momentum slices, pointed out the logarithmic divergence, and to make it vividly physical, described this as,
``You see, each momentum scale contributes its penny's worth, but at the end of the day the sum
is more than the U.S. national debt."


His advice to me as a graduate student, when I asked him whether I should work
on a programming problem (I had some prior experience with computers, and of course he was known for
his interest in them), was eminently sensible.
He said, ``You shouldn't choose a problem on the basis of the tool.
You start by thinking about the physics problem, and the computational method should be a
tool like any other. Maybe you'll solve it using computer techniques, maybe using a contour integral;
but it's very important to approach it starting from the physics because otherwise you  get lost in the use of the tool, and lose track of where you're trying to go.'' He said he'd seen lots of people get in over their heads in the computing and never get back to the physics.\footnote{A decade later, I did get involved in a computer database project, and did get in over my head.}

I took that advice to heart and didn't do anything particularly computational 
while I was his graduate student.  But when I was reading through things that he'd written, 
I wasn't completely convinced he'd been true to the spirit of his own advice.
When asked why he worked on the Kondo problem,  his answer was \cite{dibner3}:
\begin{quote}
``It comes from my utter astonishment at the capabilities of the
Hewlett-Packard pocket calculator, the one that does exponents and cosines.
And I buy this thing and I can't take my eyes off it and I have to figure out something
that I can actually do that would somehow enable me to have fun with
this calculator $\ldots$ What happened was that I worked out a very
simple version of a very compressed version of the Kondo problem,
which I could run on a pocket calculator. And then I realize that this
was something I could set up with a serious calculation on a big
computer to be quantitatively accurate."
\end{quote}
So we now know the origin of his inspiration for the problem.\footnote{In Aspen (2013), Elihu Abrahams told me, with some sheepishness, about having given a talk about the Kondo problem at Princeton in the early 70s. Afterwards, 
Ken said to him, ``I have some ideas about this problem, would you like to work on it together?''}

\medskip
Also mentioned this morning was his 1971 article \cite{prd71}.
When I asked him in '79 where he'd written down the naturalness\footnote{Starting in the late 70s, there was increased discussion of the ``naturalness problem'': the still-unresolved question of why a light scalar particle (the Higgs) is available at ``low'' energies to spontaneously break the electroweak symmetry. It is one of the many theoretical constructs that crystallized from his work, and remains with us, as part of the impetus behind supersymmetric model building and recent (so far unsuccessful) searches for supersymmetric partners to known particles.}
idea, he replied with a smile, ``Oh, that was in my paper with all possible theories of
the strong interaction, $\ldots$ except the correct one,''\footnote{In \cite{nobel},
he wrote, ``I should have anticipated the idea of asymptotic freedom but did not do so."
In 2002 \cite{dibner3}, he said in addition, ``I was among the last people to climb on board the quark idea $\ldots$ Then I developed the presupposition $\ldots$ that the Beta function would always have the wrong sign.
There's a paper that I wrote in 1971 about renormalization group and field theory,
and I discuss various alternatives for the Beta function, [and] omit
the one case that turns out to be correct. I don't mention asymptotic freedom
as a possibility because I hadn't worked on gauge theories,
and I just took for granted that the Beta functions would have the other sign.''

In 2010 \cite{videos}, he goes a bit further to say that he was so
convinced that the gauge theory Beta function would have the other sign, he simply assumed that the
deep inelastic scaling results were some intermediate phenomenon, and
that some other behavior would take over at much higher energies.}
and pointed me to the relevant paragraphs. The actual words in this article \cite{prd71} are as incisively clear as ever:
\begin{quote}
``It is interesting to note that there are no
weakly coupled scalar particles in nature; scalar particles are the
only kind of free particles whose mass term does not break either an
internal or a gauge symmetry.''
\end{quote}
In my early student days, people told me that Ken Wilson's articles
were oh-so-difficult to read, and that nobody understood them at the time.
The measure of how much he changed the research community
was that little more than five or ten years later when I read that article,
we were so thoroughly trained in the new paradigm, it read easily and beautifully:
he set up the quadratic mass divergence, and then gave a blindingly clear description of the problem that we're still grappling with.
And it's no longer easy to imagine what people were confused about by his articles from that period.

In person, he elaborated to me about setting the values of bare parameters (whether or not this constitutes more physical terms I'm not sure), and said,
``Suppose you're throwing darts at some cosmic dartboard. You're just not going to hit with 1 part
in $10^{38}$ precision if you're trying to get a 1 Gev particle starting from the Planck scale.''
This way of thinking about physics became embedded in me,
and was something that I always appreciated coming directly from him.

\begin{figure}
\centering
\includegraphics[width=.6\textwidth]{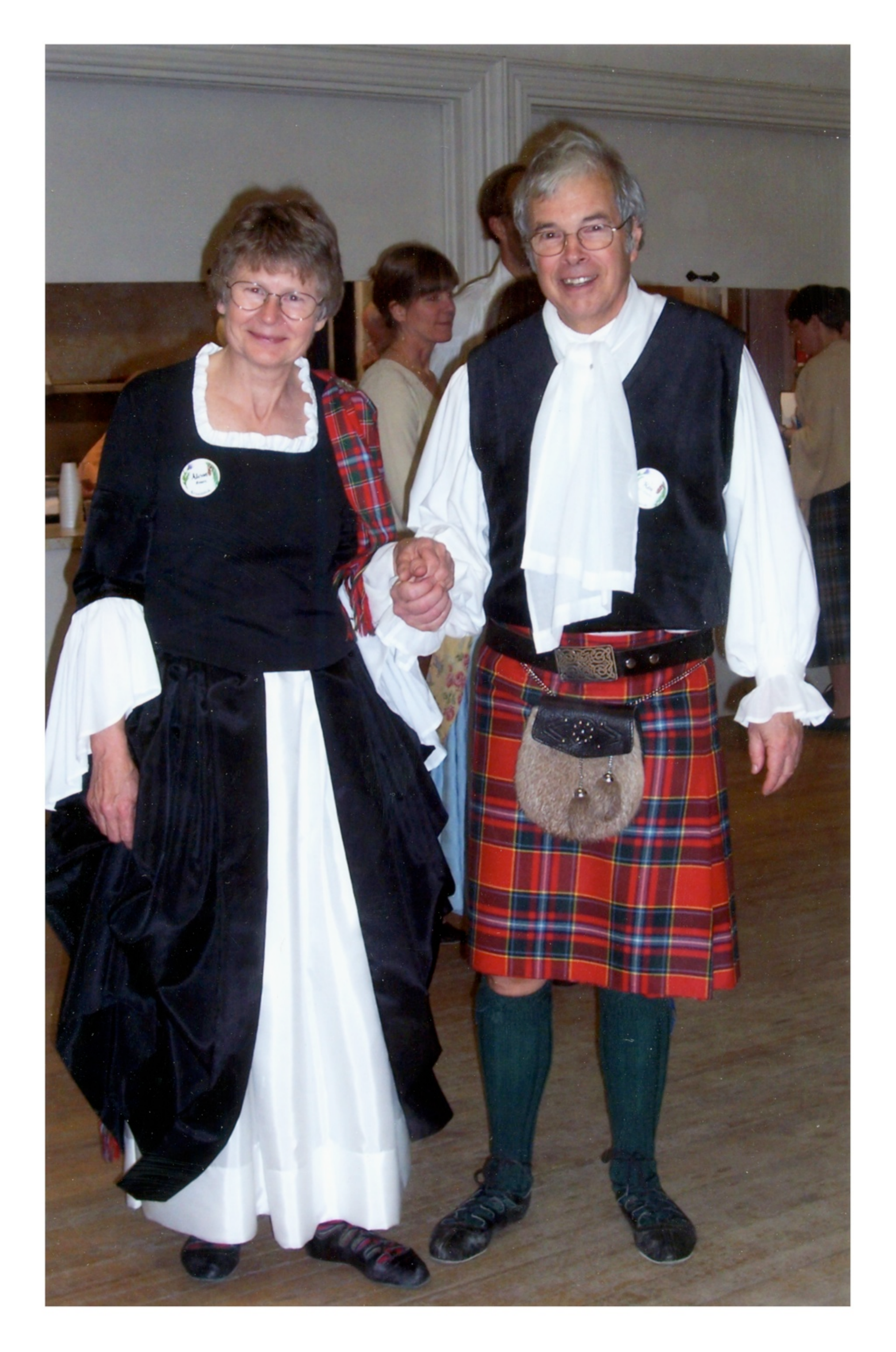}
\caption{Ken with Alison. 
This photo is here only because during dinner David Wallace asked for a photo of Ken in a kilt, so in real time I added it to the slides.}
\label{fig:6}
\end{figure}

One final comment about my graduate student experience: after Ken had looked at the first draft of my thesis,
he said to me, slowly and carefully ({\it perhaps\/} not specifically referring to the thesis):
\begin{quote}
``I was once given the advice when writing that you should go back, find all of your favorite sentences, 
{\it and delete them\/}.''
\end{quote}
I continued to convey that to others, perhaps not necessarily regarding what they had written either, as advice given to me by my thesis advisor.
It wasn't until roughly 25 years later that I mentioned the comment to Steve Strogatz,
who informed me,  ``But that's famous literary advice from  Quiller-Couch.''
Arthur Quiller-Couch, a well-known British author and Professor of English at Cambridge,
wrote a style book called ``On the Art of Writing'' (1916) \cite{QC}, and the actual quotation is:
\begin{quote}
``Whenever you feel an impulse to perpetrate a piece of exceptionally
fine writing, obey it --- wholeheartedly --- and delete it before sending
your manuscript to press. {\it Murder your darlings.\/}''
\end{quote}
So it was additionally fun to realize so much later that I'd long been propagating 
Ken's literal (i.e., physicist) transcription of this classic stylistic counsel (Fig.~\ref{fig:6}).

\section{Eighties and beyond}
\label{eighties}

Four years after getting my degree, I returned here in 
1985 to give a seminar, and by then Ken was no longer at his office in Newman
lab,\footnote{During my time at Cornell, he'd had an office on the 2nd floor  among the experimentalists,
then later moved to the 3rd floor and joined the theorists.}
having moved to the nascent Theory Center he'd founded.
I remember walking into his office, unannounced, 
he looked up and I couldn't tell if he was focusing on me or not.
And then sort of like his father, without preamble, he said,
``As you know, we've decided  we don't have enough enough computational power
to extract experimental numbers from lattice QCD."
(I still don't know if he said that to anyone and everyone who walked unexpectedly into his office.)

During that period of the 1980s,  he continued his proselytizing for 
the standardization of computer use (which we heard a lot about  this morning).
It was one of the many things that had very strongly influenced me as a graduate student: hearing him complain
that physicists should be able to travel from one institution
to another, then sit down and log in directly to their home institution, or anywhere else, and get immediately to work without learning a new operating system.\footnote{In the pre-Apple/pre-PC days, it was typical for computer companies to introduce a new operating system for each new computer. 
Adopting a more open standard, such as Unix, together with compatible networking protocols, offered the possibility of a uniform  interface for access to remote resources.
For the Cornell high energy physics group in the 80s, Peter Lepage recalls that Ken insisted to run Unix on the DEC VAX, IBM PC's and Sun workstations they'd purchased.}
He very firmly had on his mind the notion of ubiquitious network connectivity, 
frequently spoke of the need for arrays of commodity processors to leapfrog Moore's law,\footnote{The factor of two increase in processing speed every few years he said was not fast enough for many problems of interest. In this respect, he foresaw the development of commodity Linux clusters that became commonplace by the late 1980s.}
 and also was promoting better designed programming languages for parallel computing.\footnote{In the late 70s, he used to optimize code for lattice gauge theory simulations, running on a Floating Point Systems array processor, directly in assembly language. High quality optimizing compilers for parallelizing Fortran did not yet exist.
In a physics colloquium back at Harvard in 1982, he highlighted the Gibbs project, an attempt to produce a new scientific programming language.  He and Alison came to dinner with me that night at the Society of Fellows, and continued a memorable discussion.}
For this reason, it came as no surprise to me when sometime in the mid '90s, George
Strawn, who was among those responsible for shepherding the NSF-end implementation of the NSFnet in the mid 80s,\footnote{George Strawn later became NSF CIO (Chief Information Officer). When contacted in summer 2014, he added that Larry Smarr had also played an important role.} told me that on the 1982 taskforce \cite{nsfcomm} that resulted in the NSFNet recommendation, Ken was among those who argued for using the TCP/IP  (Transmission Control Protocol/Internet Protocol)
---  the basic communication language behind the thing we now call the Internet.
Apparently in the early 80s, other scientists on the committee wanted to use
DECnet\footnote{DEC was the Digital Equipment Corporation, which produced the PDP-8, PDP-10, PDP-11, and later VAX lines of computers, popular among physicists in the 70s and 80s. 
DECnet was built into the VAX/VMS operating system, and used for both remote connectivity and e-mail by the early 80s. The company went defunct when  bought  in 1988 by Compaq, which then merged with Hewlett Packard in 2002.}
because that's what they were familiar with,
and Ken had the vision to see that we'd be better off moving 
in the more open Unix direction. Strawn said Ken's voice was very influential,
and contributed to our having the current Internet.

George Strawn  told me something else back then that I found
 fascinating, and will pose as a question to the audience here.
He told me from his direct contemporary experience that there were two people in congress, one in the Senate,
and one in the House of Representatives, who were absolutely essential to getting the
Internet through congress and funded. 
Can anyone guess who those two people were? 
(Someone guesses `Al Gore', everyone laughs $\ldots$).
Al Gore, Jr. is correct! He was the one who championed it in the Senate.
And who in the House? 
(Various guesses, no one comes close.)
In the House, he said it was: Newt Gingrich!
Maybe it seems crazy, but then when you think about it, you remember
he's always interested in  off-the-wall things,
and in promoting science and technology, so it actually makes sense in some respect.

Ken also shared with us his intuition for how ubiquitous computing would become.
Peter Lepage mentioned this morning that as soon as he found out about
the networking possibilities, he anticipated physicists using it for their social
lives --- though I don't know that he anticipated the rest of the planet adopting it for their
social lives as well.

And here is an email from Michael Peskin to John Cardy (apparently intercepted by the NSA) \cite{cardy}:
\begin{quote}
``In 1976, he gave us a lecture about how, one day, we would  all be
sitting on the beach with personal computers that ran UNIX, and using
them to play games that involved exploration in three dimensions. In
1976, that seemed like a dream.''
\end{quote}

Now I'm NOT trying to say that  in 1976 he knew that Steve Jobs was about
to found Apple computer (with Wozniak), that Jobs was going to be booted from Apple 
in 1985, start his own computer company NeXT that would have
a new operating system based on Unix,  release a Unix-based computer in 1990, 
in 1996 Jobs would come back to Apple, then evolve that operating system into Mac OS X 
which would then evolve into iOS and then be installed on the iPhone, and 
become the device people use on the beach when they're playing their 3d games.
But I wouldn't be surprised if he $\ldots$


\begin{figure}
\includegraphics[width=\textwidth]{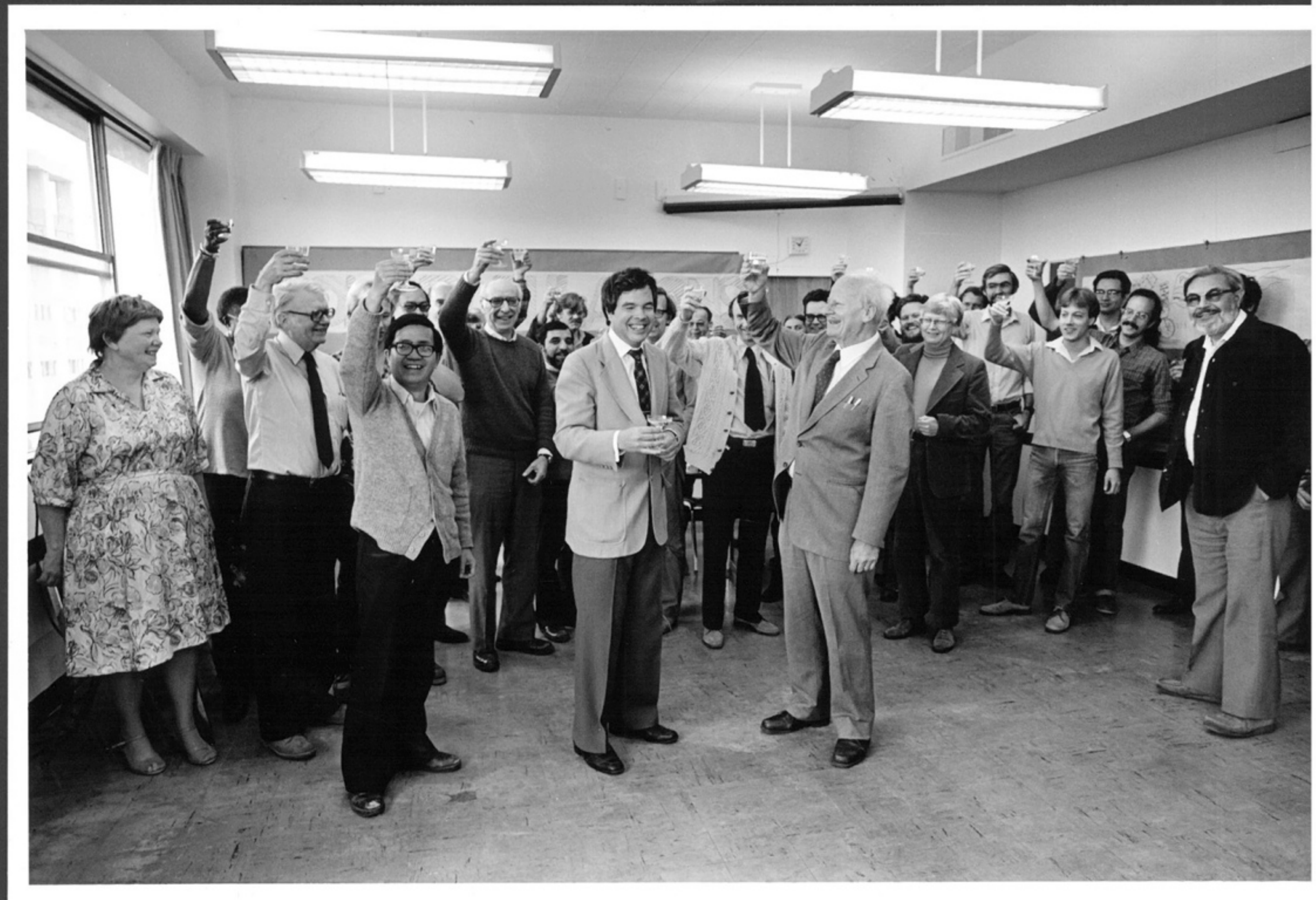}
\caption{Toasted for 1982 Nobel prize by Hans Bethe}
\label{fig:7} 
\end{figure}

I have a couple of other  photo comments.
This photo (Fig.~\ref{fig:7}) of a Nobel celebration in Newman Lab at Cornell was taken a year after I first left, but happily there are a few people I still recognize.
There's this hippy guy standing back here in the right corner, and here in the mid-left is an upper-head I can recognize just from the eyes.
I'm also very happy to see Velma Ray at the far left
(loud cheers from small subset of audience) who many of us here (Serge R., Gyan B., Michael P., Steve S., $\ldots$) will never forget. Velma was Hans Bethe's secretary, an integral part of our Newman experience, and the person who typed all of our theses in the pre-TeX era.

\begin{figure}
\includegraphics[width=\textwidth]{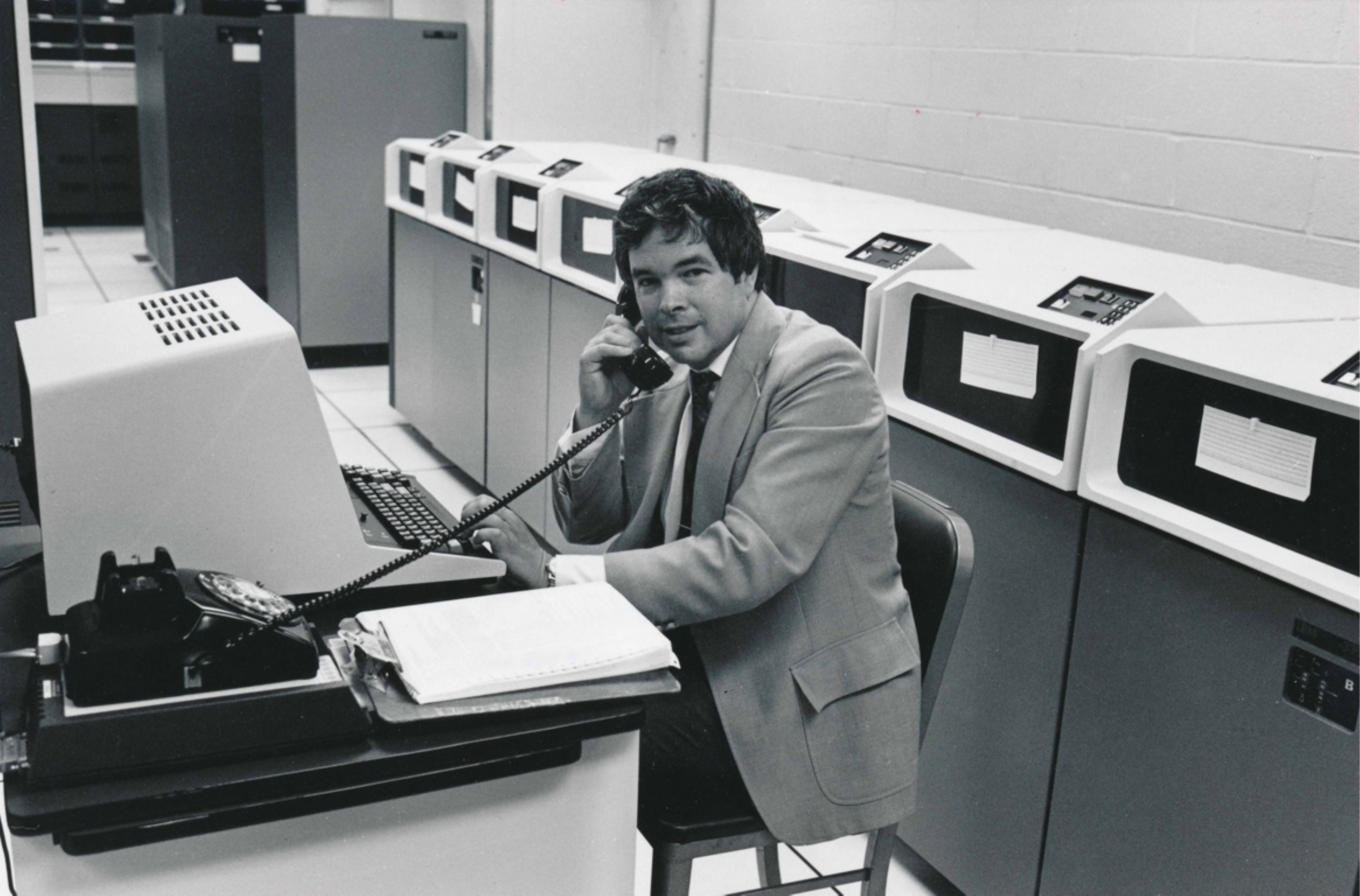}
\caption{``Why are these two separate devices?"}
\label{fig:8} 
\end{figure}

Finally here's a photo (Fig.~\ref{fig:8}) I recommended for the conference poster, 
basically encapsulating Ken as we knew him at that age, even though clearly posed.
(It's not as though the photographer happened to be present when Ken was computing away,
received a phone call, picked up the phone, and started talking.)
I recall for some of you that the large device on the desk is a physical object, known as a VT100, as opposed to the name of a terminal emulator that runs in your windowing system.
When I look at Ken on the phone with his hand on keyboard, in my mind he's thinking ``Why are these two separate devices?"

\rem{
was always very generous with refs
1982 nobel prize lecture has 147 refs, including a long historical section
even to the extent of saying "need even the ideas that didn't work:"

"I do not think the present practice of honoring only the developments
that played a key role in actual practice is really fair to all
parties, nor does it provide adequate incentives to scientists to
explore a variety of directions that could be important."

}

\section{Video Transcripts}
\label{videos}


I'm going to let Ken himself have the last word tonight.
For this, I have to give a special thanks to Stan Glazek (who's here).
He pointed me to a number of illuminating videos \cite{videos}
from a meeting Ken had with students in Poland in 2010, and also some of him
giving lectures at a school in 1994.
Last night, I watched about 2.5 hours of these videos (which was pure fun, and much
shorter than 2.5 hours because audio on computer is sped up by clipping just the pauses, so you can listen at normal pitch and not miss anything).
From this exercise, I picked out about 12 minutes of excerpts \cite{excerpts} to finish up here.\footnote{It is worth
watching the twelve minutes of on-line excerpts \cite{excerpts} transcripted here, to hear the stories directly from Ken as the audience did.}
There are many comments Ken makes about his early life and graduate student 
experiences that few of us ever heard from him, so I'm hoping this will be novel to most of you.
We're kicking ourselves here for not having invited him to come back and talk to us over the last ten years
--- you never realize until it's too late $\ldots$ But it's marvelous, after hearing so much about him all day,
we can now give him the last word.

The first video excerpt is from 1994, just to illustrate to everyone his lecture style.
The final clip  is very short, in response to a student who asked, ``When is the appropriate time
to stop taking classes, and when should one start doing research?"
And Ken's answer will be a perfect place for us to stop.

\bigskip\bigskip
 
0. KGW: ``But I want to start by reminding you that in a certain sense the
discovery of the formal rules of QCD represented a step backwards. (smile)''
\medskip

1. KGW: ``When I was a child, what I loved to do was mathematics and not much
else, other than some climbing hills $\ldots$,  paddling kayak $\ldots$, but I did
have a wonderful project in 8th grade at school, which was making a
steam engine, which is probably as close as I came to doing serious
science until I came to college.''
\medskip

2. KGW: ``My earliest memories are being fascinated by the numbers on the steam
engines that came every day to Woods Hole, which is near Boston but on
what's called Cape Cod, and I kept records of the numbers of every
steam engine coming in.''
\medskip

3. KGW: ``But by fifth grade, I was developing `serious' mathematical skills.
Namely I was taught to take cube roots of numbers by my grandfather,
and I was so fascinated by the algorithm that I used to practice it in
my head waiting for the bus going to school.''
\medskip

4. KGW: ``There was a national mathematics competition for students in college,\footnote{The Putnam Competition}
and I started participating in that math competition as a freshman and
managed to do well enough that I got invited to a dinner by all the
mathematics faculty at Harvard, as part of reward for doing well in the
competition. And that was an amazing experience, since I learned that
mathematics professors were nothing like my father, who was a
chemistry professor.''

SG: ``What was the main difference?''

KGW: ``The mathematicians were crazy (audience laughter) $\ldots$
they were wonderful at mathematics but they didn't seem to me to be
all that wonderful at real life, whereas my father understood real
life, he understood chemistry, he understood physics.
In fact he published a book which he published around 1950 which he called
`An Introduction to Scientific Research'.''\footnote{See \cite{ebw}.}
\medskip

5. KGW: ``I knew after the 8th grade that I was going to do physics, and I knew
the reason. I was going to do physics because I was certain that would
give me more interesting problems to solve than if I'd studied mathematics,
and nothing that happened in my college mathematics major\footnote{During the first video from \cite{videos}, between excerpts \#3 and \#4 above,
he explained,
``I was only majoring in mathemematics because they had a thesis requirement for mathematics and I wanted to do a thesis, and physics didn't have any equivalent." Regarding its subject matter, he said, ``I had done some research on propagation of sound underwater, in summer work at Woods Hole, and I believe I continued that work for my math thesis, but  as far as I know I have no copy $\ldots$ and no memories of what was in it."
As influential undergraduate experiences, he mentioned taking a sophomore year math course taught by George Mackey, and a course in American intellectual history taught by Arthur Schlesinger, Jr.}
changed my mind.''
\medskip

6. KGW: ``I go to graduate school in physics, and I take the first course in quantum field theory,
and I'm totally disgusted with the way it's related.
They're discussing something called renormalization group,
and it's a set of recipes, and I'm supposed to accept that these recipes work
--- no way.
I made a resolution, I would learn to do the problems that they assigned, I would learn how to turn in answers that they would accept, holding my nose all the time, and someday I was going to understand what was really going on.
And it took me ten years, but through the renormalization group\footnote{
In \cite{dibner3}, Ken said: ``In retrospect, I probably made a mistake in giving it the same name.
I probably should have given a name to distinguish the approach with one coupling and the approach with infinite couplings $\ldots$ so that $\ldots$ Gell-Mann--Low was
renormalization group A and my work was renormalization group B. We
would have gotten away from the arguments about everything reducing to
perturbative renormalization group theory.''}
work I finally convinced myself that there was a reasonable explanation for what was taught in that course .''
\medskip

7. KGW: ``I met with Murray Gell-Mann to get a thesis problem, and he suggested
a problem.\footnote{The problem was to solve the 3d Ising model, see  sec.~\ref{caltech}.}
And I thought there was no way I was working on that. So he
had to come up with another one, and it didn't take him very long.
The thing that impresses me most about Murray Gell-Mann when I think back on it,
is how quickly he got things done. I have never met anybody who could do things
as fast as he could. And so it didn't take him very long to come up
with another problem for me to work on. And so he asked me to think about
an equation, which was called the Gell-Mann--Low equation, and it was an equation which was of interest to him for what he would call low energy phenomena.
The first thing I did with that equation was study it in the limit of high energies, the exact opposite limit to anything he was interested in.
As I said, it was a small problem, but it led over ten years to larger and larger problems and to some of the key publications, in fact including the publication that was cited for the Nobel prize. But I did it by working my way from
small to large, not from starting at the large from the beginning.''\footnote{Just as the renormalization group itself works $\ldots$}
\medskip

8. KGW: ``There was a year at Caltech when Murray Gell-Mann was off in Paris,
and so Feynman was in charge of a seminar which met every week.
And most of the time he would come in and say `what should we talk about today?'
And there was one seminar when he writes that question and sort of no
decent conversation got started, and eventually he noticed that I was
talking with my neighbor. He said, `What are you talking about?'
and I said, `We're talking about a 16th century mathematics theorem
which happens to be called ``Wilson's theorem."$\,$'  And the next thing I
knew I was up -- he had me up at the blackboard explaining what this
theorem was and how you proved it.
And the theorem is that if $n$ is a prime number, and only if $n$ is a
prime number, then $n!+1$ is divisible by $n$.''\footnote{He meant $(n-1)!+1$, and later in the video corrected that, when asked how to prove it. 
``Wilson's theorem'' (stated by John Wilson and Edward Waring in 1770, earlier by Ibn al-Haytham c.\ 1000 AD, and first proved by Lagrange in 1771 \cite{wt}) is usually stated in the form:
$(n-1)! \equiv -1$ (mod$\,n$) iff $n$ is prime. To prove it, first notice that if $n$ is not prime, then
$(n-1)!\equiv0$ (mod$\,n$), since its factors occur in the product.
If $n$ is prime, then in modern language we would note that the positive integers less than $n$ form a group ($G_n$) under multiplication mod$\,n$.
Each element has an inverse distinct from itself, except for $n-1$, which is its own inverse, so 
$(n-1)!\equiv n-1\equiv -1$ (mod$\,n$).}
\medskip

9. KGW: ``In a discussion with other students, we were talking with Feynman
and one of the students asked Feynman,
``Do you have anything that you noticed about really exceptional physicists,
that characterized exceptional physicists and not normal physicists?'
And he said, `Yes, the thing that characterized them was persistence.'
That they wouldn't give up, it didn't matter how long it took.
And this was in the 1960's when we had this discussion.''
\medskip

10. KGW: ``When I turned in my thesis, which was based on this problem,
Gell-Mann was off in Paris, so Feynman was the person who had to read my thesis.
So I'll now tell you an anecdote, that after Feynman had read my thesis,
it was customary to give a seminar on one's thesis. So I gave a seminar, and
Feynman is there. And in the middle of it, another faculty member raises his
hand and says, `I find your discussion interesting, but what good is it?',
which I had no answer for.
And Feynman pipes up, and this is with an English colloquialism which I'll explain,
he says, `Don't look a gift horse in the mouth.'$\,$"
\medskip

11. KGW: ``I learned a lot that came from Gell-Mann, but not from interacting with him
personally. He was the inventor of the concepts of the renormalization group
in a paper that was published by him and Francis Low.
But to learn about that, I didn't learn anything about that from Gell-Mann
directly. I learned it from a textbook written by Bogoliubov and Shirkov,
which had a chapter on the renormalization group.
One of the other very important things I learned from Gell-Mann I didn't learn
until very recently. I was at a celebration of his career, and he
remarked that he had learned from Vicki Weisko§ pf that you had to ruthlessly
simplify the physics of a problem that you were working or learning on.''   
\medskip

12. KGW: ``Sort of the postscript to this story is thirteen years later, I
found there was a problem in condensed matter physics very similar to
the Gell-Mann--Low equation, but more complicated, but not so complicated
that I couldn't program it up and find worthwhile results on the
computers --- which were called supercomputers at that time --- but which now
can't even compete with an iPhone.''
\medskip

13. KGW: ``$\ldots$ and then figuring out what happens at about double the scale,
and then figuring out what happens on double scale again.  It's
complicated to develop the actual procedures, but there's a Scientific
American article from the late 1970s which tries to present this kind
of thinking to a lay audience. I can't quote you the reference
off-hand but anyone who looks up in the index of Scientific American\footnote{See \cite{sciam}.}
from them can find out where this article appears. And I'll say only
one more thing about that article: Scientific American wanted me to
write it, which is the way they usually do it, and then they assign an
editor, and I told them `nothing doing.'  And finally they agreed to
an arrangement where one of their editors would write the article and
I would edit it. That was a very interesting experience.''
\medskip

14. KGW: ``The U.S.\ Congress in 1990 $\ldots$''

SG: ``You are filmed, you know it.''

KGW: ``What?''

SG: ``You are on the film, so if you criticize the U.S.\ Congress $\ldots$''

KGW: ``I'm not $\ldots$ No, this is an amazing congressman, he was incredibly smart
(audience laughter),
named George Brown,\footnote{
George Edward Brown, Jr., Democratic congressman from California, was 
chairman of the Committee on Science, Space and Technology from 1991--1995,
and known as a champion for science \cite{geb}.} and he was a friend of science, and he wrote an
article in which he said he had found advice from scientists, from
physicists in particular, extremely valuable, because they would get
interested in a topic, and they just wanted to understand it.''
\medskip

15. KGW: ``My opinion is very simple: you start research at birth, and you
should never stop.''

\newpage

\begin{acknowledgements}
I thank the organizers of the symposium, Jeevak Parpia, Csaba Csaki, and Jim Sethna, for inviting me to give the after-dinner remarks, and Alison Brown for the photos used in Figs.~\ref{fig:2}, \ref{fig:3}, and \ref{fig:6}.
\end{acknowledgements}



\end{document}